\documentclass[prb,a4paper,12pt,onecolumn,superscriptaddress,amsmath,amsfonts,amssymb,preprintnumbers,citeautoscript]{revtex4}

\pdfoutput=1
\usepackage[T1]{fontenc}\usepackage[latin1]{inputenc}
\usepackage{graphicx,booktabs,microtype,afterpage,sidecap, times, soul}
\usepackage[charter]{mathdesign}

\usepackage[colorlinks,plainpages=false,linkcolor=black,urlcolor=blue,citecolor=black,pdfpagemode=UseNone,pdfstartview=FitBH]{hyperref}

\usepackage{graphicx}   
\usepackage{verbatim}   
\usepackage{color}      
\usepackage{subfigure}
\usepackage{gensymb}
\usepackage{textcomp}
\usepackage{hyperref}   
\usepackage{longtable}
\usepackage{csquotes}
\usepackage{commath}
\usepackage[english]{babel} 
\usepackage{multirow} 
\usepackage{cleveref} 
\usepackage[normalem]{ulem} 
\usepackage{xcolor} 
\usepackage{bm}
\usepackage{color}
\usepackage{epstopdf}

\makeatletter
\renewcommand{\@evenfoot}{\hfill\raisebox{-1em}{\bf\thepage}\hfill}
\renewcommand{\@oddfoot}{\hfill\raisebox{-1em}{\bf\thepage}\hfill}
\makeatother

\newcommand{\sio}{\ensuremath{\mathrm{SrIrO_3}}}
\newcommand{\dso}{\ensuremath{\mathrm{DyScO_3}}}

\newcommand{\eg}{$\rm{E_{g}}$}
\newcommand{\ttwo}{$\rm{T_{2g}}$}
\newcommand{\cpp}{$\chi^{{\prime}{\prime}}$}
\newcommand{\cppr}{$R\chi^{{\prime}{\prime}}$}

\newcommand{\cppe}{$\chi^{{\prime}{\prime}}_{e}$}

\newcommand{\icm}{\,$\rm{cm^{-1}}$ }
\newcommand\redsout{\bgroup\markoverwith{\textcolor{red}{\rule[0.5ex]{2pt}{0.4pt}}}\ULon}

\graphicspath{{//}}

\begin{document}

\title{\large{\textbf{ Strange semimetal   dynamics in $\mathbf{SrIrO_3}$ }}}

\author{K. Sen}
\affiliation{Institut f\"{u}r Quantenmaterialien und -technologien, Karlsruher Institut f\"{u}r Technologie, 76021 Karlsruhe, Germany\\}

\author{D. Fuchs}
\affiliation{Institut f\"{u}r Quantenmaterialien und -technologien, Karlsruher Institut f\"{u}r Technologie, 76021 Karlsruhe, Germany\\}

\author{R. Heid}
\affiliation{Institut f\"{u}r Quantenmaterialien und -technologien, Karlsruher Institut f\"{u}r Technologie, 76021 Karlsruhe, Germany\\}

\author{K. Kleindienst}
\affiliation{Institut f\"{u}r Quantenmaterialien und -technologien, Karlsruher Institut f\"{u}r Technologie, 76021 Karlsruhe, Germany\\}

\author{K. Wolff}
\affiliation{Institut f\"{u}r Quantenmaterialien und -technologien, Karlsruher Institut f\"{u}r Technologie, 76021 Karlsruhe, Germany\\}
\author{J. Schmalian}
\affiliation{Institut f\"{u}r Quantenmaterialien und -technologien, Karlsruher Institut f\"{u}r Technologie, 76021 Karlsruhe, Germany\\}
\affiliation{Institut f\"{u}r Theorie der Kondensierten Materie, Karlsruher Institut f\"{u}r Technologie, 76131 Karlsruhe, Germany \\}

\author{M. Le Tacon}
\affiliation{Institut f\"{u}r Quantenmaterialien und -technologien, Karlsruher Institut f\"{u}r Technologie, 76021 Karlsruhe, Germany\\}
\date{\today}



\begin{abstract}
\center\bigskip\thispagestyle{plain}
\begin{minipage}{\textwidth}\textbf{The interplay of electronic correlations, multi-orbital excitations, and strong spin-orbit coupling is a fertile ground for new states of matter in quantum materials. Here, we report on a confocal Raman scattering study of momentum-resolved charge dynamics from a thin film of semimetallic perovskite $\mathbf{SrIrO_3}$. We demonstrate that the charge dynamics, characterized by a broad continuum, is well described in terms of the marginal Fermi liquid phenomenology. In addition, over a wide temperature regime,  the inverse scattering time is for all momenta  close to the Planckian limit $\mathbf{\tau^{-1}_{\hbar}=k_{\rm B} T/\hbar}$. Thus,  $\mathbf{SrIrO_3}$ is a semimetallic multi-band system that is as correlated as, for example, the cuprate superconductors.
The usual challenge to resolve the charge dynamics in multi-band systems with very different mobilities is circumvented by taking advantage of the momentum space selectivity of polarized electronic Raman scattering.
The Raman responses of both hole- and electron-pockets display an electronic continuum extending far beyond 1000\icm ($\sim$125 meV), much larger than allowed by the phase space for creating particle-hole pairs in a regular Fermi liquid. Analyzing this response in the framework of a memory function formalism, we are able to extract the frequency dependent scattering rate and mass enhancement factor of both types of charge carriers, which in turn allows us to determine the carrier-dependent mobilities and electrical resistivities. The results are well consistent with transport measurement and demonstrate the potential of this approach to investigate the charge dynamics in multi-band systems. }\end{minipage}
\end{abstract}

\maketitle\thispagestyle{empty}\clearpage
\section*{Introduction}

Numerous quantum materials have exotic electronic properties that cannot be accounted for within the canonical framework of the Fermi liquid theory. They attract increasing attention both because of the profound challenge they pose to our fundamental understanding of electrons in condensed matter~\cite{Keimer_NaturePhysics2017} and because of their technological potential~\cite{Tokura_NaturePhysics2017}. Over the last decade, the spin-orbit coupling (SOC), that describes the strength of the interaction between the spin and the orbital motion of a quasiparticle, has been identified as one of the major ingredient for the realization of novel quantum phases of matter~\cite{Witczak-Krempa_ARCM2014}, encompassing in particular topological or axion insulators, quantum spin liquids as well as Dirac or Weyl topological semimetals. The interplay between electronic correlations and semimetal behavior has for example been discussed in the context of quantum criticality in Dirac systems\cite{Sheehy_PRL2007,Assaad_PRX2013} and topological and non-Fermi-liquid states that emerge near quadratic band touching points\cite{Moon_PRL2013}.

5$d$ transition metal oxides are particularly interesting materials in this respect. In the iridate perovskites from the Ruddlesden-Popper series $\rm{Sr_{n+1}Ir_{n}O_{3n+1}}$, the crystal field lifts the degeneracy of the 5d levels of the octahedrally coordinated Ir$^{4+}$ ions. Combined with electron-electron correlations (U ${\sim}2$\,eV) and large spin-orbit coupling (${\sim}0.4$\,eV), this yields a peculiar type of insulating antiferromagnetic phases at half-filling in $\rm{Sr_{2}IrO_{4}}$ (n=1) and $\rm{Sr_{3}Ir_{2}O_{7}}$ (n=2) compounds. This phase is known as the spin-orbital Mott state~\cite{Kim_Science2009}, in which $J_{\rm eff}$ = 1/2 pseudo-spins rather than pure spins order magnetically and is widely seen as a novel platform for unconventional superconducting states.
Additionally, a plethora of fascinating exotic physical phenomena, encompassing pseudogap or Fermi arc states, have been discovered in these systems and remain to be understood~\cite{Bertinshaw_ARCM2019} . 
The n = $\rm{\infty}$ phase of this series, {\sio} (SIO), exhibits, on the other hand, a semimetallic and paramagnetic ground state~\cite{Zhao_JAP2008, Cao_PRB2007,Fujioka_PRB2017}, and has long been predicted to host Dirac quasiparticles near the Fermi energy $E_F$~\cite{Zeb_PRB2012,Carter_PRB2013}, making it a potential realization of a correlated Dirac semimetal~\cite{Chen_NatCom2015}. {\par}

Recent Angle-Resolved Photoemission Spectroscopy (ARPES) studies on SIO directly confirmed the semimetallic character of this compound, in which both a heavy hole-like and a lighter electron-like band cross the Fermi level $E_F$~\cite{Nie_PRL2015, Liu_SciRep2016}, as predicted by first principles calculations~\cite{Zeb_PRB2012,Carter_PRB2013, Zhang_PRL2013}. Despite a very steep and quasi-linear dispersion of the electron-like band, the theoretically predicted~\cite{Zeb_PRB2012, Carter_PRB2013} symmetry protected degeneracy of the Dirac point has been found to be lifted~\cite{ Liu_SciRep2016}. Although evidences for the proximity of SIO to a magnetic insulating state have been reported~\cite{Biswas_JAP2014, Matsuno_PRL2015, Zhang_Review2018}, the sizeable mixing of the $J_{\rm eff}$ = 1/2 and 3/2 character of the narrow bands crossing $E_F$ strongly contrasts with the case of $\rm{Sr_{2}IrO_{4}}$.

Significant efforts have been devoted to understand the charge dynamics in this system. However, disentangling the contribution of both types of carriers in such multi-band systems is generally a delicate task using conventional transport methods. In the particular case of SIO, since the effective mass of quasi-linearly dispersing electrons is much smaller than the one of the heavier holes~\cite{Nie_PRL2015}, the electrons have a higher mobility ($\mu={e\tau_0}/{m^*}$, $\tau_0$: static relaxation time) and dominate Hall effect measurements~\cite{Kleindienst_PRB2018, Fujioka_PRB2017, Zhang_PRB2015}. Another study suggests that the electrical conductivity and thermopower are affected by both, electron- and hole-like carriers~\cite{Manca_PRB2018}. Both types of charge carriers further contribute to the optical conductivity~\cite{Moon_PRL2008, Fujioka_PRB2017, Fujioka_JPSJ2018}. 

Consequently, and to the best of our knowledge, neither the static nor the dynamical scattering rates and mass enhancements of the electron- and hole-like carriers were determined experimentally in SIO. Investigations of the charge dynamics in this system are further complicated by the fact that, unlike the n=1 and 2 compounds from which single crystals can be grown, the perovskite phase of SIO is metastable and can only be synthesised in polycrystalline or in thin film forms. In the latter case, for most of the substrates, significant lattice mismatches induce large strain effects, as well as \textit{e.g.} twin or grain boundaries that considerably affect the electro-dynamics of the system. Indeed, the Hall coefficient, the carrier density or the resistivity have recently been reported to be strongly substrate and film-thickness dependent~\cite{Zhang_PRB2015, Biswas_JAP2014, Groenendijk_PRL2017,Liu_PRM2017}, indicating strong strain-relaxation effects. {\par}

As both first-principles calculations~\cite{Zeb_PRB2012, Carter_PRB2013} and ARPES~\cite{Nie_PRL2015, Liu_SciRep2016} investigations have shown that the hole- and electron-pockets crossing the Fermi level occupy distinct regions of the reciprocal space, the way towards momentum-selective studies of the charge dynamics~\cite{Devereaux_RMP07} in SIO via electronic Raman scattering (ERS) is paved. In this communication, we take full advantage of the polarization selection rules of ERS to independently investigate the intrinsic charge dynamics of the hole- and electron-like carriers in a fully strain-relaxed 50 nm thick SIO thin film. Exploiting the confocality of our micro-Raman set-up, we have been able to extract the electronic response from the thin film and to reveal the existence, for both types of charge carriers, of a flat electronic continuum extending at least up to 1000 cm$^{-1}$ ($\sim$ 125 meV). This is strongly reminiscent of the Raman response of doped high-$T_c$ superconducting cuprates~\cite{Bozovic_PRL1987, Cooper_PRB88, Virosztek_PRB1992} . This in turn suggests that such continuum might be a universal feature of correlated electron system on the verge of a Mott transition. The reported electronic continuum is a characteristic feature of the marginal Fermi liquid (MFL) phenomenology and has been analyzed using a memory function formalism, which allowed us in particular to quantitatively extract charge-carrier-resolved frequency-dependent scattering rates $\Gamma(\omega,T)$ and mass enhancements $ m^{*}/m_{b}=1+\lambda(\omega,T)$, where $m^{*}$ and $m_{b}$ are the effective mass renormalized by e-e interactions and the bare band mass, respectively. For both type of carriers, these quantities amount for a rather broad temperature regime to an inverse quasiparticle time 
\begin{equation}
\hbar \tau^{-1}=\Gamma /(1+\lambda)
\label{tauqp}
\end{equation}
 that is surprisingly  close to the Planckian limit\cite{Sachdev_1999, Zaanen_Nature2004} $\tau^{-1}_{\hbar}=k_{\rm B}T /\hbar$, even though at lowest $T$ the MFL theory yields $\tau^{-1}\propto T/\log(D/T)$, where D is an appropriate cut-off frequency (typically the bandwidth of the relevant degrees of freedom).  More generally, this work demonstrates the potential of polarized electronic Raman scattering for the study of scattering rate, mass enhancement and mobility of charge carriers in semimetals and other multiband systems.
{\par}

\section*{Experiment}
\paragraph*{Samples.} Epitaxial thin films of {\sio} (50\,nm) were grown by pulsed laser deposition on orthorhombic ($Pnma$) ($101$)-oriented $\rm{DyScO_3}$ (DSO) substrates and characterized, as described in Ref.~\onlinecite{Kleindienst_PRB2018}. As shown in Fig.~\ref{fig-polsel}a, our films are $(101)$-oriented, with their \textit{b-axis} parallel to the one of DSO substrate. The tilt pattern of the $\rm{IrO_6}$ octahedra, $a^-b^+a^-$ in Glazer notation is the same as that of bulk {\sio}.{\par}

\paragraph*{Fermi Surface.} We determined the Fermi surface of the (101)-oriented SIO with respect to its pseudocubic unit cell (UC), in which the Ir-atoms sit at the corners of the pseudocube using first-principles calculations, as depicted in Fig.~\ref{fig-polsel}b. In agreement with previous reports and ARPES studies~\cite{Nie_PRL2015, Liu_SciRep2016}, we find multiple electron bands crossing the Fermi Level around $(\pm\pi/2, \pm\pi/2)$, and several flat hole bands extending along the $\Gamma$-X  directions from the $\Gamma$ and Y points. 

\paragraph*{Polarization selection rules.} 
Confocal Raman scattering experiments were carried out in backscattering geometry. As detailed in the Method section (and in the supplementary information Figs. S1 and S2), the confocal geometry allows us to accurately separate the intrinsic Raman response of the SIO thin film from that of the DSO substrate~\cite{Hepting_PRL2014, Hepting_PhysicaB2015}. Fig.~\ref{fig-polsel}c summarizes the Raman polarization selection rules relevant for the present study, which allow to selectively probe the charge dynamics on different sections of the Fermi surface of SIO. 
The $\rm{XYZ}$-coordinate system used to label the polarization of the light is oriented along the Ir-Ir bonds of the pseudo-cubic unit cell (incident laser propagates in the direction perpendicular to the plane of the film, along the $\rm{Z}$-axis). The Raman vertex will be accordingly described within the cubic $\rm{O_h}$ point group. When the incident photon polarization is at $45^{\circ}$ to the Ir-O-Ir bonds and the scattered photons are detected in a crossed-polarization configuration (Porto's notation: $\rm{Z(X'Y')\bar{Z}}$), the corresponding electronic Raman structure factor has the {\eg} symmetry ($x^2-y^2$), which probes the dynamics of charge carriers with momentum along the $(0,\pm\pi)$ and $(\pm\pi,0)$ directions of the 1-Ir Brillouin zone (BZ), as shown in Fig.~\ref{fig-polsel}b. 
Similarly, the electronic Raman structure factor in $\rm{Z(XY)\bar{Z}}$ has the symmetry of {\ttwo} ($xy$), which probes the charge carriers whose momentum is located around the  $(\pm\pi/2, \pm\pi/2)$ in 1-Ir BZ, as shown in Fig.~\ref{fig-polsel}c. 
Thus, in summary, $\rm{X'Y'}$ and $\rm{XY}$ geometries predominantly probe dynamics of hole- and electron-pockets, respectively. 

\section*{Results and discussion}
\begin{figure*}[!htbp] 
\centering
\includegraphics[width=0.8\columnwidth]{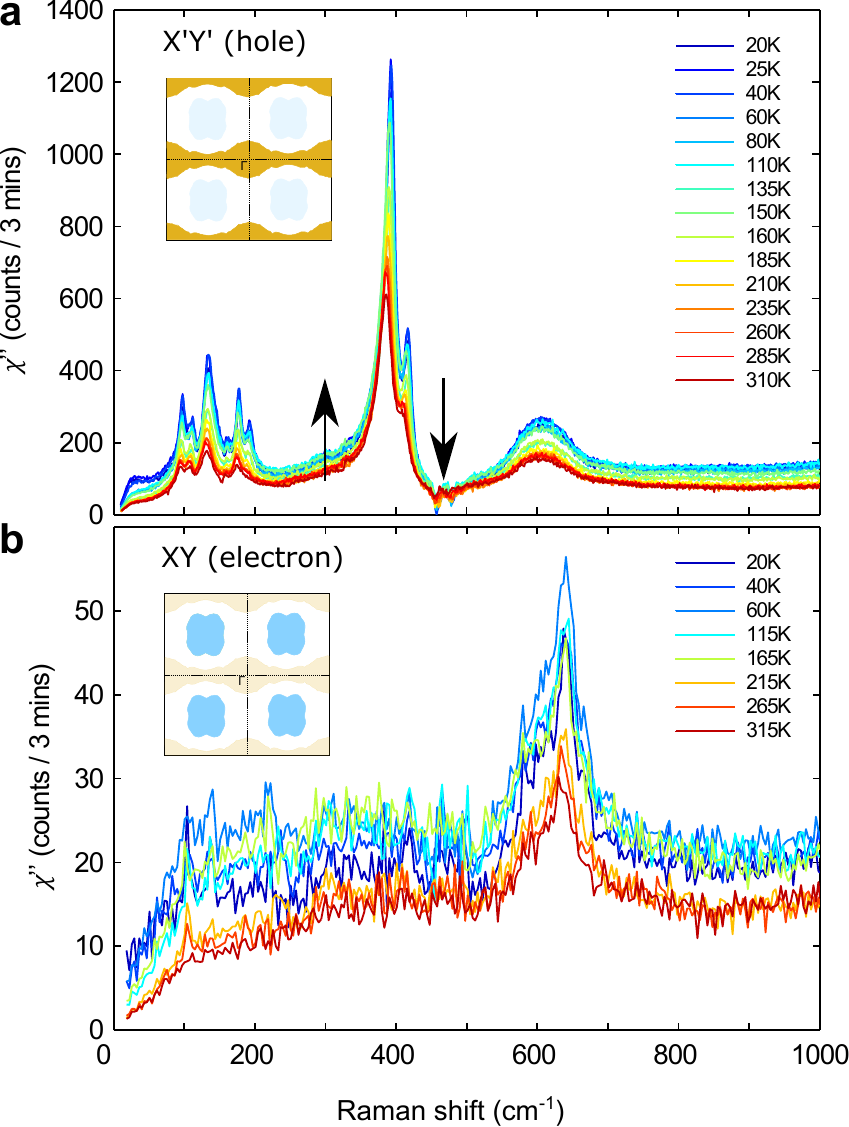}
\caption{\label{fig-allT-data} {\textbf{Symmetry-dependent Raman spectra.}} (\textbf{a}) Raman spectra ({\cppr}$(\omega)$) in $\rm{X'Y'}$ at several temperatures. The arrows indicate spectral weight loss and gain above and below the phonon modes around $400$\,cm$^{-1}$, respectively. (\textbf{b}) The corresponding plot in $\rm{XY}$ at a few temperatures.} 
\end{figure*}

\paragraph*{Raman responses in $\rm{X'Y'}$ and $\rm{XY}$.} Figs.~\ref{fig-allT-data}a and~\ref{fig-allT-data}b display the Raman response {\cpp}$(\omega)$ of the SIO film in $\rm{X'Y'}$ and $\rm{XY}$ geometries, respectively, at several temperatures ($T$). In both cases, {\cpp}$(\omega)$ consists of sharp phonons that are superimposed to a flat continuum extending over the whole range of investigated frequencies.
Before discussing the origin of this continuum, it is crucial to check that it does not arise from an artefact related to the subtraction of the substrate's contribution. To do this, we applied the same procedure to a 60\,nm thick film of insulating $\rm{CeO_2}$ (see Supplementary Fig. S3), in which no continuum in the Raman response is expected. Its absence in the experimental spectrum therefore confirms that the featureless Raman response observed in SIO is intrinsic. This continuum is strongly reminiscent of the electronic Raman response encountered in high-$T_c$ cuprates~\cite{Bozovic_PRL1987, Cooper_PRB88, Virosztek_PRB1992, Devereaux_RMP07} or Fe-based superconductors~\cite{Kretzschmar_NatPhys2016, Gallais_PRL2013},
in which it arises from the creation of particle-hole excitations across the Fermi level ($E_F$)~\cite{Devereaux_RMP07}.
Our interpretation for SIO is supported by the observation that the continuum is $\sim$5 times bigger in $\rm{X'Y'}$ than in $\rm{XY}$, as qualitatively expected from the larger density of states (DOS) of the flat hole-like bands at $(0,\pm\pi)$ and $(\pm\pi,0)$ in comparison to the almost linearly dispersing electron-like bands around $(\pm\pi/2,\pm\pi/2)$. 
{\par}

\paragraph*{Phonon contribution.}
In order to analyze the $\rm{X'Y'}$ Raman continuum in details, it is necessary to subtract the other spectral features recorded (Fig.~\ref{fig-allT-data}a). We observe 9 peaks (7 between 90 and 220\icm, and 2 more around 400\icm ) that sharpen and harden as temperature is lowered and that correspond to Raman active lattice vibrations of SIO. Their detailed analysis will be the subject of a separate study. The broader peak centred around 600\,cm$^{-1}$ is only weakly temperature dependent, and its frequency is larger than the calculated cut-off frequency of the phonon spectrum. We therefore tentatively assess this to a double phonon process and its contribution to the Raman response can be easily subtracted. A close inspection of the two most intense modes around 400\icm reveals that the lower energy one is significantly asymmetric, \textit{i.e.} spectral-weight gain toward low-frequency seemingly occurs at a cost of spectral-weight loss at high-frequency (highlighted by arrows in Fig.~\ref{fig-allT-data}a). This lineshape, known as a Fano resonance~\cite{Fano_PR1961}, generally arises from the coupling between a discrete excitation and a continuum. Crucially here, this implies that the electronic continuum cannot be analyzed independently and that its coupling to the phonon should explicitly be taken into account (see Methods). The contribution of all the other phonon lines can be modelled using regular Lorentzian lineshapes. The analysis in the $\rm{XY}$ channel is comparatively simple, since the corresponding spectra have only a broad feature above 600\icm, which we also attribute to double-phonon scattering. It can be easily subtracted out and that does not distort the underlying $e$-continuum scattering background. {\par}

\paragraph*{Electronic Raman Scattering.}
To extract the pure electronic contribution {\cppe}$(\omega,T)$ from the $\rm{X'Y'}$ data, we follow the approach described in Ref.~\onlinecite{Chen_PRB1993} and use a phenomenological model (see ~\cref{eq_globalfit} in Methods) to fit the Raman response {\cpp}$(\omega,T)$. Importantly, this approach (described in more details in Methods) requires an analytical form for the {\cppe}$(\omega,T)$ that we express in terms of a memory function $M(\omega,T)$. The method was originally introduced by G\"{o}tze and W\"{o}lfle to calculate the frequency-dependent optical conductivity of metals~\cite{Gotze_PRB1972}, and has been generalised to extract dynamical scattering rate $\Gamma(\omega,T)$ and mass enhancement factor $1+\lambda(\omega, T)$ of charge carriers. It was adapted to analyze ERS in high-$T_c$ cuprates~\cite{Opel_PRB2000} and Fe-based superconductors~\cite{Kretzschmar_NatPhys2016}.
In the memory function formalism, the experimentally challenging determination of the ERS intensity in absolute units is not required and has not been attempted here. 
The memory function parametrisation of the total electronic Raman response function reads\cite{Opel_PRB2000}:
\begin{equation}\label{eq_totalRaman}
\chi_e (\omega,T)=\frac{M(\omega,T)}{\hbar\omega+M(\omega,T)},
\end{equation}
where $M(\omega,T)$ is the memory function. {\par}
The real ($\chi^{\prime}_e$) and imaginary ({\cppe}) parts of $\chi_e$ are even and odd functions of the Raman shift $\omega$, respectively. Thus we can write
\begin{equation}\label{eq_memory}
M(\omega,T)=\hbar\omega\lambda(\omega,T)+i\Gamma(\omega,T),
\end{equation}
where, $\lambda(\omega,T)$ and $\Gamma(\omega,T)$ are real even functions  that are related 
via a Kramers-Kronig relation.

 
  {\cppe}$(\omega,T)$ follows from~\cref{eq_totalRaman,eq_memory} as 
\begin{equation}\label{eq_imchi}
\chi^{\prime\prime}_e(\omega,T)=\frac{\hbar\omega\Gamma(\omega,T)}{\left[\hbar\omega(1+\lambda(\omega,T))\right]^2+\left[\Gamma(\omega,T)\right]^2}.
\end{equation}
In analogy to the optical conductivity, we can introduce the quasiparticle scattering time $\tau$ of eq.\ref{tauqp}, that includes the mass renormalization, contained in the real part of the memory function,  \textit{i.e.} in $\lambda$, see \textit{e.g.} Ref.\,\onlinecite{Bruin_Science2013}.
 
To proceed, we analyze specific forms of the scattering rate $\Gamma(\omega,T)$ and determine $\lambda(\omega,T)$ via Kramers-Kronig transformation.  $\Gamma(\omega,T)$  essentially contains two terms, arising from 
temperature independent impurity scattering ($\Gamma_{imp}$) and electron-electron scattering respectively. 
For a conventional metal (Fermi liquid), this electron-electron scattering rate is given by:
\begin{equation} \label{eq_Gamma_FL}
\Gamma_{FL}(\omega,T)=\frac{g}{E_F}\left[\left(\hbar\omega\right)^2+\left(\beta {k_B}{T}\right)^2\right],
\end{equation}
where the dimensionless coupling constant $g$  characterizes the correlation strength~\cite{Tytarenko_SciRep2015}.   $\beta$ is a numerical coefficient that determines the relative importance of thermal versus dynamic excitations for the scattering rate. In the case of a single particle scattering rate holds $\beta=\pi$ while a quantum Boltzmann analysis for the optical conductivity, a two-particle quantity, yields $\beta=2\pi$.\cite{Gurzhi_JETP1959, Chubukov_PRB2012} The analysis of Refs.~\onlinecite{Maslov_PRB2012,Stricker_PRL2014} revealed that, in principle, $\pi \leq \beta < \infty$ is possible, depending on the relative strength of elastic and inelastic scattering processes. This was for instance reported for $\rm{URu_2Si_2}$~\cite{Nagel_PNAS2012} and $\rm{UPt_3}$~\cite{Sulewski_PRB1988}.
 


The dynamics of both types of charge carriers inferred from our data can not be adequately described in terms of this Fermi-liquid expression (see supplementary information).
This yield our first important conclusion, namely charge dynamics in SIO is non-Fermi liquid like. To gain further insights, we therefore modelled the Raman response in terms of a singular quasiparticles scattering rate as $\Gamma(\omega,T) \propto (\left(\hbar\omega\right)^2+({\beta}{k_B}T)^2)^{\alpha}$, with
an exponent $\alpha \neq$ 1. Acceptable fits of our data can be achieved with values for $\alpha$ in the range  $0.45 \leq \alpha \leq 0.6$ (see Supplementary Note 2). For simplicity we chose $\alpha_{\rm MFL} = 1/2$ which corresponds to the marginal Fermi liquid (MFL) phenomenology~\cite{Varma_PRL1989}. This choice is also motivated by the fact that  the observed featureless ERS continuum of SIO is strongly reminiscent to that of the high-$T_c$ cuprates~\cite{Bozovic_PRL1987, Cooper_PRB88, Virosztek_PRB1992}, in which it is considered as one of the hallmark of the MFL phenomenology~\cite{Varma_PRL1989, Virosztek_PRB1992}. 
Crucially, the dependence of the different fitting parameters on the exact value of  $\alpha$ lies well within the corresponding error bars. Choosing $\alpha$ in the vicinity of $\alpha_{\rm MFL} $ will merely change some logarithmic dependencies to power laws with small exponents.

\begin{figure*}[!htbp] 
\centering
\includegraphics[width=0.5\columnwidth]{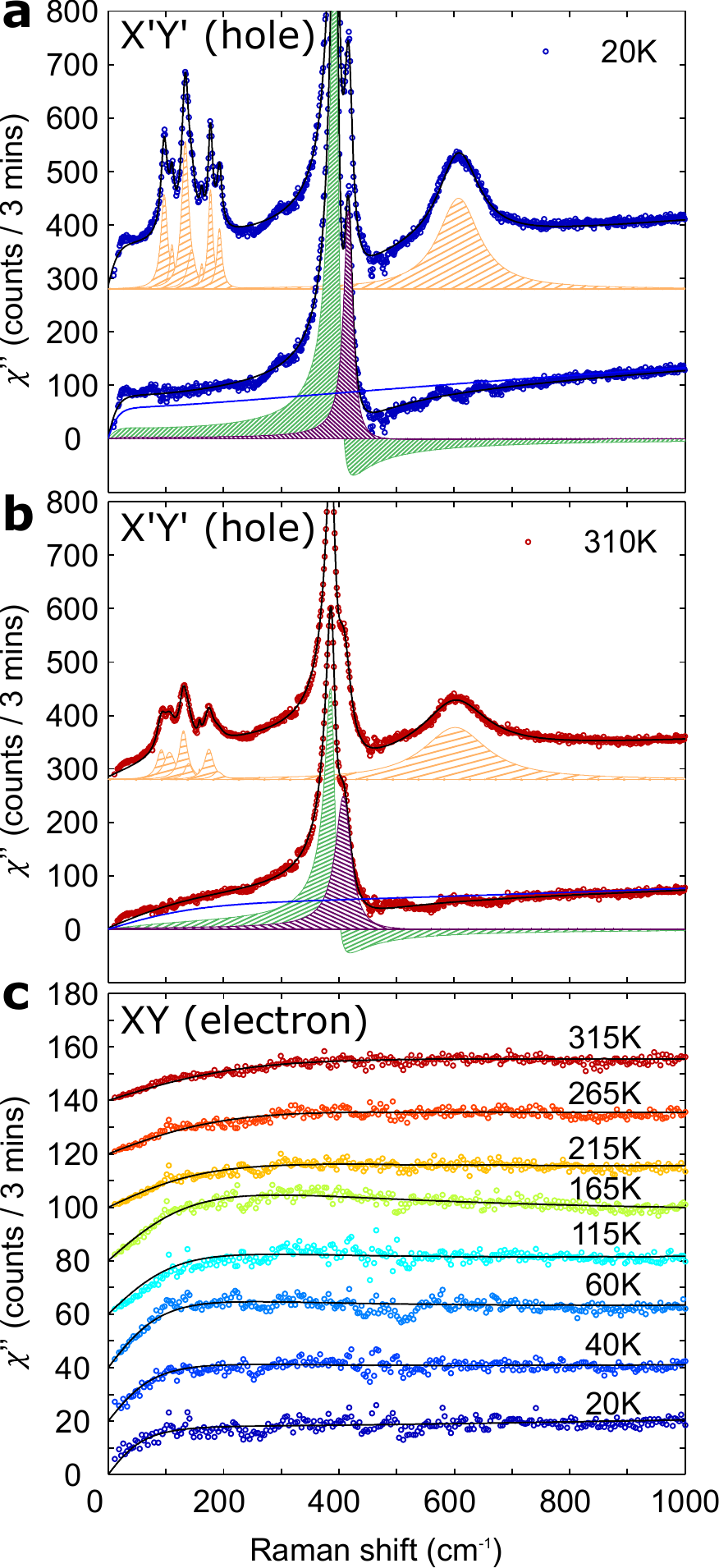}
\caption{\label{fig-globalfit} {\textbf{Fits with MFL model to extract electronic Raman scattering response.}} (\textbf{a}) Raman spectrum at $20$\,K in $\rm{X'Y'}$ (blue circles). Top: a representative fit to the Raman spectrum at $20$\,K with a phenomenological model (solid line in black). Phonon modes of Lorentzian lineshape are shaded in orange.  Bottom: The Raman spectrum after subtraction of the Lorentzian modes. One of the remaining phonon modes manifest Fano-like asymmetric feature (shaded in green and purple). The solid line in blue is the best-suited $e$-continuum scattering according to the marginal Fermi-liquid (MFL) model in ~\cref{eq_imchi}. (\textbf{b}) The corresponding Raman spectrum and fit at $T=310$\,K. (\textbf{c}) Raman spectra ({\cppr}$(\omega)$) in $\rm{XY}$ (symbols) at several temperatures. The data were vertically shifted for representation. The solid lines are the best-suited $e$-continuum scattering backgrounds according to the MFL model.}
\end{figure*}

\paragraph*{Marginal Fermi liquid phenomenology.}
The characteristics of the MFL is that the quasiparticle weight in the single particle spectrum vanishes like $1/\log\left(\frac{D}{\hbar \omega}\right)$, i.e.
the spectrum tends to be  weakly incoherent near the Fermi energy~\cite{Varma_PRL1989}. 
According to Refs.~\onlinecite{Varma_PRL1989,Virosztek_PRB1992}, the scattering rate of the Raman response $\Gamma(\omega,T)$ for a MFL can be written as
\begin{equation} \label{eq_GammaMFL}
\Gamma_{\rm MFL}(\omega,T)=g \sqrt{\left(\hbar\omega\right)^2+({\beta}{k_B}T)^2},
\end{equation}
where $g$ is  again a dimensionless strength of the coupling. Microscopically it is determined by the ratio of the effective electron-electron ($e$-$e$) correlations, \textit{i.e.} onsite Coulomb repulsion ($U$) and the   bandwidth ($W$).   $\beta$ plays an analogous role as for the Fermi liquid rate discussed above.  
{\par}
The total scattering rate is then given by:
\begin{equation} \label{eq_GammaTotal}
\Gamma(\omega,T)=\left[\Gamma_{\rm MFL}(\omega,T)+\Gamma_{\rm imp}\right]\phi({\omega}/D),
\end{equation}
where $\phi$ is an appropriate cut-off function with a cut-off frequency of $D$. Since ${\hbar}D$ is in the order of the bandwidth, its value for SIO is several hundreds of meV. Thereby, $\hbar{D}$ only weakly affects $\Gamma(\omega,T)$ in the investigated range of frequencies  (see Supplementary Note 2). More details regarding the memory function formalism based on the MFL model are given in the Supplementary Note 2. 
Representative resulting global fits (including phonon lines and their eventual coupling to the continuum) to {\cpp}$(\omega)$ at $T=20$ and 310\,K for  $\rm{X'Y'}$ are shown in~Figs.~\ref{fig-globalfit}a and ~\ref{fig-globalfit}b, respectively. The agreement with the data both at low and high temperatures is excellent and validates the MFL approach to describe the charge dynamics for the hole-like carriers in SIO. We note that the Fano asymmetry parameter for the 391\,cm$^{-1}$ mode is essentially temperature-independent in the investigated range, indicating no significant change in the electronic structure at the Fermi surface as a function of temperature. Similarly, the  data recorded in the $\rm{XY}$ channel (Fig.~\ref{fig-globalfit}c) can also be very well fitted in this framework.

\begin{figure*}[!htbp] 
\centering
\includegraphics[width=\columnwidth]{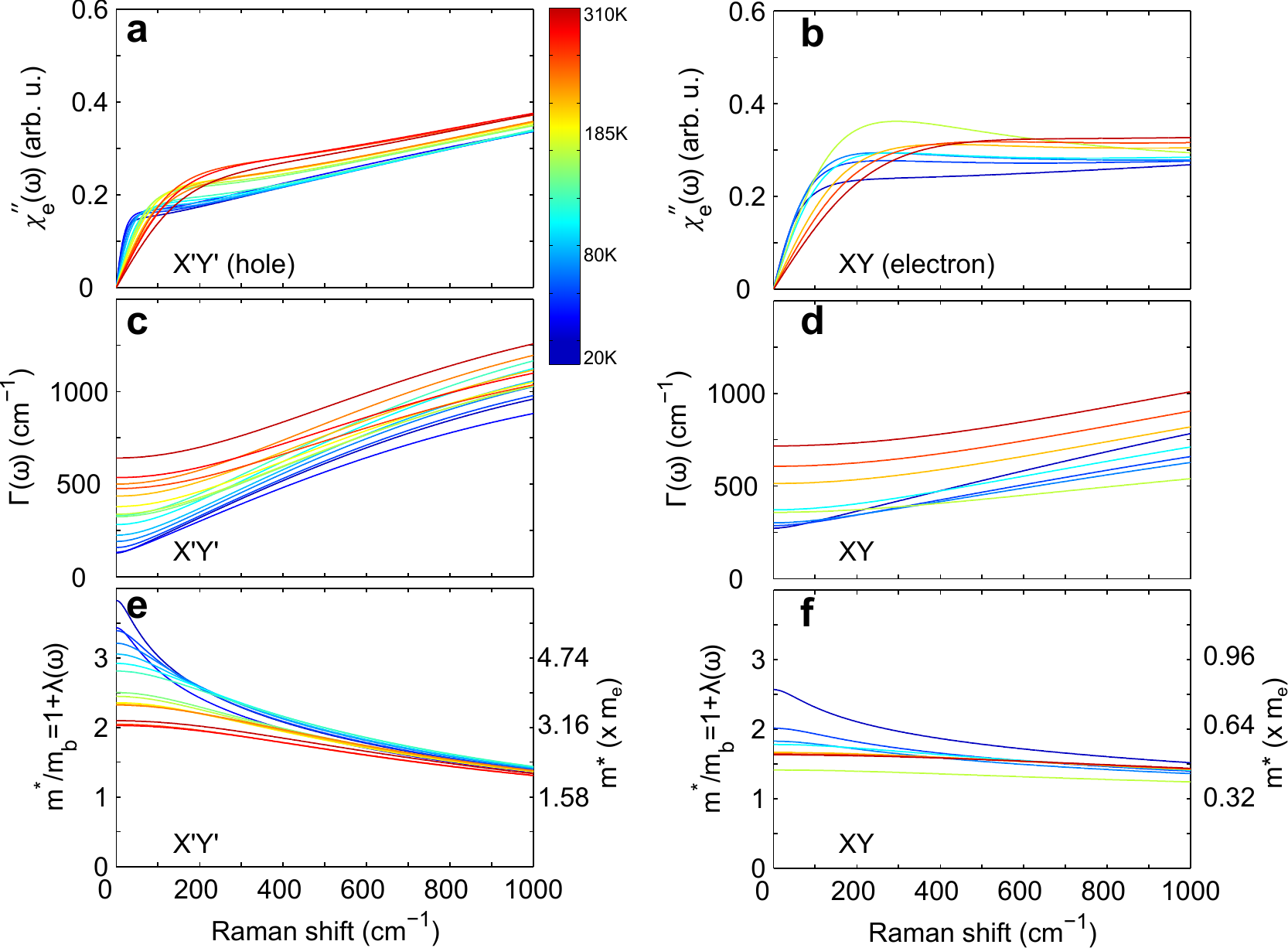}
\caption{\label{fig-econt} {\textbf{Electronic Raman scattering and charge transport parameters from the MFL fits.}} Electronic Raman scattering response ({\cppe}$(\omega)$) at several temperatures in (\textbf{a}) $\rm{X'Y'}$ and (\textbf{b}) $\rm{XY}$. Dynamical scattering rates ($\Gamma({\omega,T})$) in (\textbf{c}) $\rm{X'Y'}$ and (\textbf{d}) $\rm{XY}$. The corresponding dynamical mass enhancement factors ($m^*/m_b(\omega,T)$) in (\textbf{e}) and (\textbf{f}), respectively. They were obtained from the marginal Fermi-liquid (MFL) fits, as shown in Fig.~\ref{fig-globalfit}.} 
\end{figure*}

\paragraph*{Analysis of Raman spectra with the MFL model.} 
In Figs.~\ref{fig-econt}a and ~\ref{fig-econt}b, we display the result of our analysis, \textit{i.e.} the pure electronic Raman responses {\cppe}$(\omega)$ in $\rm{X'Y'}$ and $\rm{XY}$ channels, respectively. 
The responses in the two channels appear quite distinct at all temperatures. This is particularly striking at energies $\hbar\omega \gg k_BT$, in which we detect an increase of {\cppe}$(\omega)$ in $\rm{X'Y'}$, which, in contrast, remains essentially constant in $\rm{XY}$. 
To better understand the origin of such difference, we take a closer look at the values of the parameters $g$, $\beta$ and $\Gamma_{\rm imp}$ obtained for each Fermi pockets
from our fitting procedure. {\par}

First of all, we note that the $\Gamma_{\rm imp}$ and $\beta$ parameters are both temperature-independent quantities, but both can take very different values for the different Fermi pockets.
From our analysis, they respectively amount to $\Gamma^{\rm hole}_{\rm imp}=12.6{\pm}2$\,{\rm meV} and $\beta^{\rm hole}= 1.84{\pm}0.3$ ($\sim 0.6\pi$) in $\rm{X'Y'}$ and $\Gamma^{\rm elec}_{\rm imp}=30.6{\pm}5$\,{\rm meV} and $\beta^{\rm elec}=3.23{\pm}0.5$ ($\sim \pi$) in $\rm{XY}$.

As shown in Fig. S7a in the supplementary information, the value of the dimensionless coupling constant $g$ is only weakly temperature dependent.
It is  at least two times larger in $\rm{X'Y'}$ ($g \sim 1.2$) than in $\rm{XY}$ ($g \sim 0.5$), indicating stronger electron-electron correlations on the hole-pockets compared to the electron-pockets in SIO. 
The values for the electron pockets  are similar to those reported for superconducting cuprates $\rm{YBa_2Cu_3O_7}$ ( $g=$  0.55) and $\rm{Bi_2Sr_2CaCu_2O_8}$ ($g =$0.4)~\cite{Virosztek_PRB1992}. Thus, it seems that the hole pockets in SIO are at least as correlated as in the cuprates.

This has a direct impact on the frequency dependence of the dynamical scattering rate $\Gamma(\omega,T)$ and mass renormalization $1+\lambda(\omega, T) = m^*/m_b(\omega,T)$ of the two types of carriers. Indeed, the larger $g$ in $\rm{X'Y'}$ compared to $\rm{XY}$ results in a steeper $\Gamma(\omega,T)$ for the holes than for the electrons, as shown in Figs.~\ref{fig-econt}c and ~\ref{fig-econt}d, respectively. A larger $g$ value also enhances the effective mass of the holes more than that of the electrons (see Figs.~\ref{fig-econt}e-f), yielding an effective mass enhancement $m_0^*/m_b=1+\lambda_0 = 1+\lambda(\omega\rightarrow0, T)$ of 3.8 for the former against 2.6 for the latter at low temperatures.

\begin{figure*}[!htbp] 
\centering
\includegraphics[width=\columnwidth]{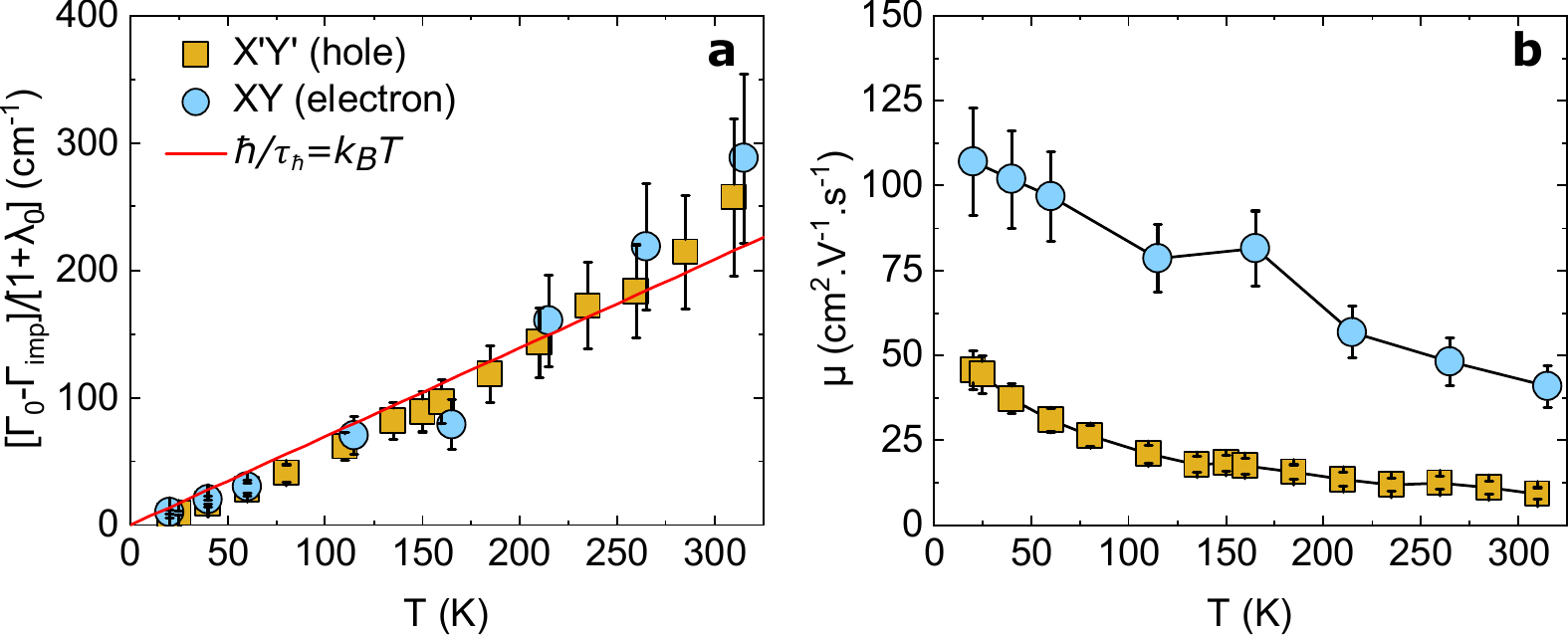}
\caption{\label{fig-static} {\textbf{Static charge transport properties.}} (\textbf{a}) Inverse inelastic quasiparticle scattering time of conduction holes and electrons as function of temperature, both  be following the universal Planckian limit ($\hbar\tau_{\hbar}^{-1}=k_BT$). (\textbf{b}) Calculated DC mobility for both charge carriers as function of temperature. The error bars reflect the maximum proportional error calculated from the errors on individual fitting parameters (see supplementary information).}
\end{figure*}

\paragraph*{Transport parameters in the static limit.} To allow direct comparison with the other experiments, we estimate transport parameters such as the resistivity and mobility of each type of charge carrier from $\Gamma(\omega,T)$ and $1+\lambda(\omega\, T)$ in the $\rm{X'Y'}$ and $\rm{XY}$ channels in the static limit ($\omega \rightarrow 0$), respectively. 
Under the  assumption that the rates for impurity and inelastic scattering are additive, we obtain for both carrier types that the inverse inelastic quasiparticle time $ \hbar \tau_{\rm inel.}^{-1}=\left(\Gamma_0-\Gamma_{\rm imp}\right)/\left(1+\lambda_0\right)$
is very close to the Planckian scattering bound\cite{Sachdev_1999, Zaanen_Nature2004}, $k_{\rm B}T$, as shown in Fig.~\ref{fig-static}a. This bound plays an important role in  the discussion of transport quantities\cite{Bruin_Science2013} where it serves as a measure of the degree of correlation strength of a quantum material. The subsequent comparison of $\tau^{-1}$  with transport parameters strongly suggests that the rate obtained by Raman scattering allows for the same conclusion.
Our observation that $\tau_{\rm inel.}\sim \tau_{\hbar}$ for holes and electrons, characterized by distinct coupling constants $g$, cut-off frequencies $D$, and coefficients $\beta$, is remarkable. It cannot be the ultimate low-$T$ limit of a MFL.  At lowest temperatures the logarithmic growth of the MFL  mass renormalization yields $\hbar \tau^{-1}\sim \frac{\pi\beta}{2}\frac{k_{B}T}{\log\frac{D}{\beta T}}$, where
 holes and electrons reach different quasiparticle rates.  The quantitative values of the mass renormalization $\lambda(T)<3$ (see supplementary information) 	suggests that we have not yet reached this low-$T$ regime.  At this point it is unclear whether the observed "universality" of $\tau$ for different momenta is a coincidence or whether there is a deeper underlying principle at work.


The knowledge of the static quasiparticle scattering time $\tau_0 = \hbar (1+\lambda_0)/\Gamma_0$, together with the assumption that it behaves similar to the one in the electric conductivity, allows for an estimate of the electron mobility $\mu=\frac{e \tau_0}{m_0^*}$  ($e$ is the electronic charge) and of the Drude-like DC resistivity $\rho_0=m_0^*/(n e^2 \tau_0)$ of each type of charge carrier. The former quantity can be directly extracted from our analysis from which we determined the temperature dependent mass enhancement (with respect to the calculated band masses $m_{b,h}=1.58m_e$ and $m_{b,e}=0.32m_e$, see methods) and static rates (Fig.~\ref{fig-econt}), while the latter requires the knowledge of the charge carrier densities, as discussed below.
We summarise the static transport parameters extracted from our analysis at room temperature and 20K in Table~\ref{tab:static}, and we plot the extracted mobility of the two charge carriers in Fig.~\ref{fig-static}b. Despite a larger value of the coupling constant $g$ for the holes compared to the electrons, the static relaxation rate  $\Gamma_0(T)={g}{\beta}k_BT+\Gamma_{imp}$ is comparable for the two types of carriers (see Fig.S7b), given the smaller value of $\beta$ and $\Gamma_{imp}$ for the holes. 

\begin{table}[htbp]
\centering
\begin{tabular}{ c c c c c c c  }
\hline\hline
  &  $m_0^*/m_b $ & $\Gamma_0$ & $\tau_0$ &$\mu$ & $\rho_0$ \\
 &   $= 1+ \lambda_0$ & (\icm) & $ ({\times}10^{-15}$\,s) & ($\rm{cm^2.V^{-1}.s^{-1}}$) &  ($\rm{m\Omega.cm}$)  \\
\hline\hline
 hole  (RT) &  2.1 & 642 & 17.3 & 9.2 & 1.4 \\
 electron (RT)  & 1.6 & 716 & 12.1 & 40.7 & 0.5  \\
  \hline
hole  (20 K) & 3.8 & 130 & 157 & 45.6 & 0.3 \\
 electron (20 K) & 2.6 & 273 & 50 & 107 & 0.2 \\
  \hline
\end{tabular}
\caption{DC mass enhancement factor ($m^*/m_b$), DC scattering rate $\Gamma_0=\Gamma(\omega \rightarrow 0)$, static quasiparticle relaxation time ($\tau_0 = \hbar (1+\lambda_0)/\Gamma_0$), mobility ($\mu$) and resistivity ($\rho_0$) of the conduction holes and electrons around 310\,K and 20\,K.}\label{tab:static}
\end{table}

The drastically different mobilities of the two types charge carriers naturally account for the electron-like linear and negative Hall resistance ($R_{XY}$ vs. $H$) measured on our sample~\cite{Kleindienst_PRB2018} and previously reported in the literature~\cite{Nie_PRL2015,Manca_PRB2018}. To detect the field-dependent non-linearities of the Hall resistance arising from the multiband nature of the system, magnetic fields much larger than 15T would be required (see supplementary note 3). This nevertheless implies that the estimate of the charge-carrier density inferred from $R_{XY}$, assuming that the Hall resistance is only caused by electron-carriers, is overestimated.
The value obtained for this sample at 2K, $n_e=3.0{\times}10^{20}$\,cm$^{-3}$ is only weakly temperature dependent (it increases by about 6\% at room temperature), and is of the same order of magnitude, albeit larger, than the estimate of ref.~\onlinecite{Manca_PRB2018} ($n_e=1.6{\times}10^{20}$\,cm$^{-3}$). Note that the sample of ref.~\onlinecite{Manca_PRB2018} was grown on a different substrate, and that the transport properties have been reported to be substrate dependent~\cite{Zhang_PRB2015}. 
Based on this, and on the assumption that $n_h \sim 1.5{\times} n_e$ (as in ref.~\onlinecite{Manca_PRB2018}), we have estimated the resistivity for each charge-carrier. The total resistivity can be calculated by assuming that the two channels conduct current as parallel resistors, yielding $\rho_0(20K) \sim$0.1$\,\rm{m\Omega.cm}$ and $\rho_0(300K) \sim$ 0.4$\rm{m\Omega.cm}$. 
Thus, the value we determine here for the room temperature resistivity of our SIO film is reasonably close to that determined using conventional resistivity measurements (see supplementary Fig. S9), and is consistent with the values of 0.5-2\,$\rm{m\Omega.cm}$ reported in the literature~\cite{Biswas_JAP2014,Nie_PRL2015,Groenendijk_PRL2017,Liu_PRM2017}. Given, as discussed above, that the Hall measurements tends to overestimate the electron charge carrier density, both values for the resistivity extracted from the ERS measurements are slightly underestimated, and the agreement with transport is overall very satisfactory at room temperature. The situation is a bit different at low temperature where our transport measurement indicate an upturn in the resistivity, which is not seen here. This upturn is associated with weak localization effects that occur over the length scale probed in the transport measurement, and to which the present experimental approach, which probes the system over a length scale of $\sim 2 \mu m$ (set by the size of our laser spot), is not sensitive. The proposed approach therefore allows us to extract the intrinsic dynamics of the quasiparticles in SIO.  {\par}

\section*{Conclusion}
To summarize, in this work we combined the selection rules of polarized Raman scattering and the high spatial resolution of confocal geometry to investigate independently the dynamics of electron- and hole-like charge carriers in a film of semimetallic {\sio} as thin as 50 nm. 
 We find that neither of them can be described within the framework of the Fermi liquid theory, and that the electronic Raman response can be well modelled using marginal Fermi liquid phenomenology. 
Using mass enhancement and the DC scattering rate obtained from this analysis allow us to retrieve the mobility for the two types of charge carriers. The results confirmed the much larger mobility of the electron carriers that generally dominate transport experiments. The proposed approach demonstrates the power of Raman scattering to resolve the dynamics of charge carriers in correlated semimetals and more generally multiband systems.

\section*{Methods}

\paragraph*{Polarization-resolved confocal Raman scattering.}
Confocal Raman scattering experiments were performed with a Jobin-Yvon LabRAM HR Evolution spectrometer in backscattering geometry. We used a He-Ne laser ($\lambda=632.8$\,nm) with a laser power of ${\leq}0.8$\,mW that was focused on the sample with a $100${$\times$} magnification long-working-distance ($7.6$\,mm) objective. The laser spot size is $\approx2$\,$\rm{{\mu}m}$ in diameter. 
The sample was placed on a motorised stage that translate along the incident beam direction which, in combination with a 50\,$\rm{{\mu}m}$ confocal hole at the entrance of the spectrometer provides the high spatial resolution required to investigate thin films. To do so, we first optimised the signal from the film by recording Raman spectra of the sample at different position across the focal point of the microscope. Reference spectra from the substrate was obtained focusing the laser deep in the substrate, and subtracted from the total response (Supplementary Note 1). {\par}
Spectra were recorded using a grating of $600$\,grooves/mm, yielding the spectrometer resolution of $1.8$\,$\rm{cm^{-1}}$. Any resonance effect in the spectral background was discarded by measuring with a second laser of $\lambda=532$\,nm. Temperature-dependent measurements were carried out by placing the samples in a He-flow Konti-cryostat. All representative spectra were corrected for Bose-factors and laser-induced heating of ${\leq}15$\,K, estimated from systematic study of Raman data with laser power.  

\paragraph*{Gobal fit.} To disentangle the electronic contribution from the total (electron+phonon) Raman response {\cppr}$(\omega)$ in $\rm{X'Y'}$ configuration, we fitted the data using the approach proposed by Chen et al.~\cite{Chen_PRB1993}. The Raman response is given by:
\begin{align}\label{eq_globalfit}
R\chi^{\prime\prime}(\omega)&=R\left[\chi^{\prime\prime}_e(\omega)+\sum_{i=1}^{2}\frac{g_i^2}{\Lambda_i(\omega)\left[1+\epsilon_i^2(\omega)\right]}\left[S_i^2(\omega)-2\epsilon_i(\omega)S_i(\omega)\chi_e^{\prime\prime}(\omega)-{\chi_e^{\prime\prime}}^2(\omega)\right]\right] \nonumber \\
&+R\left[\sum_{j=1}^{8}\frac{C_jw_j}{2\pi}\frac{1}{\left(\omega-\omega_{c,j}\right)^2+\left(\frac{w_j}{2}\right)^2}\right] 
\end{align} 
where $R$ is a proportionality constant which is a function of temperature in the present case, imaginary part ({\cppe}$(\omega)$) of the pure electronic Raman response is entangled with two phonon modes through the Fano-terms expressed in the second term of \cref{eq_globalfit}, and the last term accounts for the remaining eight phonon modes which have Lorentzian lineshapes. $\epsilon_i(\omega)$ and $S_i(\omega)$ are defined as $\epsilon_i(\omega)=({\omega^2-\Omega_i^{2}})/{2\omega_{0,i}\Lambda_i(\omega)}$ and $S_i(\omega)=S_{0,i}+\chi_e^{\prime}(\omega)$, respectively. Here, $\chi_e^{\prime}(\omega)$ is the real part of the electronic Raman response, and was obtained from \cref{eq_totalRaman}. Renormalized phonon frequency and linewidth in presence of electron-phonon coupling ($g$) are defined as $\Omega_i^2=\omega_{0,i}^2-2\omega_{0,i}g_i^2\chi^{\prime}_e(\omega)$ and $\Lambda_i(\omega)=\Lambda_{0,i}+g_i^2\chi^{\prime\prime}_e(\omega)$, respectively. $\omega_{0,i}$ and $\Lambda_{0,i}$ are the intrinsic phonon frequency and linewidth, respectively. $S_{0,i}$ combines Raman phononic matrix element ($T_{p,i}$), Raman electronic matrix element ($T_{e,i}$) and $g_i$, and it reads $S_{0,i}=T_{p,i}/(T_{e,i}{\cdot}g_i)$. $c_j$, $w_j$ and $\omega_{c,j}$ in the last term of \cref{eq_globalfit} are area, linewidth and center of the $j^{\rm{th}}$ Lorentzian phonon, respectively. The parameter that determines the Fano asymmetry reads 
\begin{equation}\label{eq_fano}
q_i(\omega)=-\frac{S_i(\omega)}{\chi^{\prime\prime}_e(\omega)},
\end{equation}
in which $1/q_i^2(\omega_{0,i})$ describes the asymmetry in the phonon lineshape. {\par}

\paragraph*{DFT Calculation} DFT calculations were performed for SrIrO$_3$ using the mixed-basis pseudopotential method~\cite{Louie,Meyer}. This method combines plane waves and local functions in the basis set, which allow an efficient description of more localized components of the valence states.
In this study we used plane waves up to a kinetic energy of 28 Ry, augmented by local functions of $s, p, d$ type at the Sr and Ir sites, respectively, and of s and p type at the O sites.
Norm-conserving pseudopotentials were constructed following the description of Vanderbilt~\cite{Vanderbilt} including the Sr-$4s$, Sr-$4p$, Ir-$5s$, Ir-$5p$, and O-$2s$ semi-core states in the valence space. The local-density approximation in the parametrization of Perdew and Wang\cite{PW} has been employed. Spin-orbit interaction is consistently incorporated in the DFT Hamiltonian by using a spinor formulation and by including spin-orbit components of the pseudopotentials~\cite{SOC}. 
For the orthorhombic structure of SrIrO$_3$ we took the experimental lattice parameters of the film on the DyScO$_3$ substrate~\cite{Kleindienst_PRB2018}. Atomic positions were relaxed until the remaining forces were below 10$^{-3}$ Ry/au. For the relaxation, it was sufficient to employ an orthorhombic 8x8x6 k-point mesh for Brillouin zone (BZ) integration in conjunction with a Gaussian smearing of 0.1 eV. Due to the presence of flat hole like bands close to the Fermi level, band structure and subsequent Fermi surfaces were determined with the tetrahedron method without any broadening, and with a very dense 32x32x24 k-point mesh.

\begin{figure*}[!htbp] 
\centering
\includegraphics[width=\columnwidth]{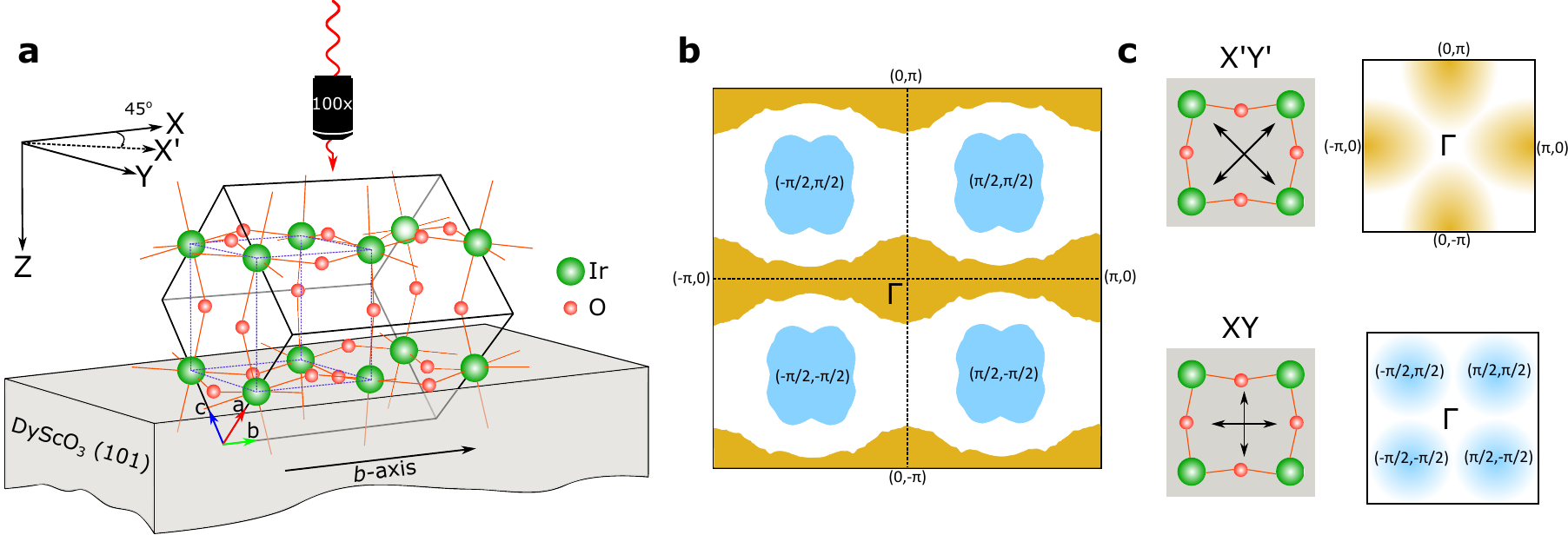}
\caption{\label{fig-polsel} {\textbf{Polarization selection rules.}} (\textbf{a}) Sketch illustrating orientation of {\sio} crystal unit cell of $Pnma$ space group on $(101)$-{\dso} substrate. A pseudocubic unit cell of {\sio} is designated by blue dashed lines. The XYZ-coordinate system concerning the backscattering Raman experiments is marked in black-arrows. Incoming laser propagates along Z-axis, as shown by a red-arrow through a microscope objective.  (\textbf{b}) Calculated Fermi surface is shown in 1-Ir BZ for the respective pseudocubic unit cell. The electron pockets are at $(\pm\pi/2,\pm\pi/2)$, whereas the rest arises from hole-like bands. (\textbf{c}) Electronic Raman scattering structure factors are shown on 1-Ir (pseudocubic unit cell) Brillouin zone (right panels) for the respective scattering configurations, namely $\rm{Z(X'Y')\bar{Z}}$ and $\rm{Z(XY)\bar{Z}}$, as depicted in left panels. The corresponding symmetries are {\eg} and {\ttwo} according to the point group of $\rm{O_h}$, respectively.}
\end{figure*}

The Fermi surface shown in Fig.~\ref{fig-polsel}b represents a top view for a (101)-orientation of SrIrO$_3$. The electron pockets at ($\pm \pi/2$,$\pm \pi/2$) in 1-Ir BZ are a robust feature associated to steep bands, whereas the hole-like Fermi surface is an open surface arising from very flat bands and is very sensitive to structural details. Rough estimates for Fermi velocities $v_F$ of holes and electrons have been derived from the band dispersion along high-symmetry directions. Effective band masses were estimated using the formula $m_b =(\hbar q_F)/v_F$ , where $q_F$ denotes the wavevector of the Fermi level crossing measured with respect to the pocket centers. This procedure gave average band masses for holes and electrons of 1.58 $m_e$ and 0.32 $m_e$, respectively.

\noindent\textbf{Acknowledgements}
We are grateful to I. Paul, Y. Gallais and R. Eder for valuable discussions. We acknowledge support by the state of Baden-W\"{u}rttemberg through bwHPC.

\noindent\textbf{Author contributions}
K.S. and M.L.T. conceived the project. K.S. acquired and analyzed the Raman scattering data under the supervision of M.L.T. {\sio} thin films were prepared and characterized by D.F., K.K. and K.W. R.H. performed first-principle calculations to obtain Fermi surface. J.S. developed the theoretical framework which was used to fit the Raman scattering data. K.S., J. S. and M.L.T. wrote the manuscript with the inputs from all the co-authors. 

\noindent\textbf{Additional Information} Correspondence and requests for materials should be addressed to
M.~L.~T. (\href{mailto:matthieu.letacon@kit.edu}{matthieu.letacon@kit.edu}).

\newpage

\bigskip

\end{document}